\documentclass[orivec]{llncs}
\pdfoutput=1

\usepackage[utf8]{inputenc}
\usepackage[english]{babel}
\usepackage[babel]{csquotes}
\usepackage{microtype}

\usepackage{amssymb}
\usepackage{mathtools}
\allowdisplaybreaks

\DeclarePairedDelimiter\floor{\lfloor}{\rfloor} 
\DeclarePairedDelimiter\ceil{\lceil}{\rceil} 
 
\DeclarePairedDelimiterX\Set[2]{\lbrace}{\rbrace}{#1\,\delimsize\vert\,\mathopen{}#2}

\newcommand{\R}{\ensuremath{\bbbr}}

\usepackage{graphicx}
\usepackage[caption=false]{subfig}
\usepackage[usenames, dvipsnames,svgnames]{xcolor}
\usepackage{float}

\usepackage[hidelinks, breaklinks]{hyperref}

\title{A Lower Bound on Supporting Predecessor~Search in \(k\) sorted Arrays\thanks{This work was presented at the Young Researcher Workshop on Automata, Languages and Programming (YR-ICALP 2015), July 5th, 2015 in Kyoto, Japan.}}
\author{Carsten Grimm\inst{1}\fnmsep\inst{2}}
\institute{Otto-von-Guericke-Universität Magdeburg, Magdeburg, Germany \and Carleton University, Ottawa, Ontario, Canada}

\begin{document}

\maketitle
 \vspace{-2\baselineskip}
\begin{abstract}
We seek to perform efficient queries for the predecessor among \(n\) values stored in \(k\) sorted arrays. Evading the \(\Omega(n \log k)\) lower bound from merging \(k\) arrays, we support predecessor queries in \(O(\log n)\) time after \(O(n \log(\frac{k}{\log n}))\) construction time. By applying Ben-Or's technique, we establish that this is optimal for strict predecessor queries, i.e., every data structure supporting \(O(\log n)\)-time strict predecessor queries requires \(\Omega(n \log(\frac{k}{\log n}))\) construction time. Our approach generalizes as a template for deriving similar lower bounds on the construction time of data structures with some desired query time. 
\end{abstract}

\section{Introduction}

We are given \(k\) sorted arrays \(A_1, A_2,\dots, A_k\) storing \(n\) values in total. Let \(A\) be the sorted array that results from merging \(A_1, A_2,\dots, A_k\). We would like to support efficient queries for the predecessor of any query value \(q\) in the array \(A\), i.e., for the largest value in \(A\) that is smaller than or equal to \(q\). However, we would like to accomplish this goal without explicitly constructing \(A\) and thereby avoiding the lower bound of \(\Omega(n \log k)\) from merging \(k\) sorted arrays.

By combining partial merging with fractional cascading~\cite{chazelle1986fractional}, we support \(O(\log n)\)-time predecessor queries in \(A\) after \(O(n \log (\frac k {\log n}))\) construction. As our main contribution, we prove that the resulting data structure is, in fact, optimal when considering strict predecessor queries, i.e., queries for the largest entry of \(A\) that is strictly smaller than the query value \(q\). By applying Ben-Or's technique~\cite{benor1983lower}, we establish a lower bound of \(\Omega(n \log (\frac{k}{\log n}))\) on the construction time of every data structure that supports strict predecessor queries in \(O(\log n)\) time. 

We are interested in lower bounds on the construction time of data structures for predecessor search in multiple arrays, because we wish to derive lower bounds on the construction time of more complex data structures.

\section{Related Work} \label{sec::relatedwork}
While lower bounds for predecessor search have been extensively studied in various models of computation, such as the cell probe model~\cite{sen2008lower}, we are unaware of any results regarding the version studied in this work. One variant of predecessor search that comes close is the setting of fractional cascading, where we seek to identify the predecessor of a query value in each array as opposed to the overall predecessor. Chazelle and Guibas~\cite{chazelle1986fractional} support simultaneous predecessor queries in \(k\) sorted arrays in \(O(k + \log n)\) time after \(O(n)\) construction. When \(k = O(\log n)\), fractional cascading solves our version of predecessor search optimally. 

Ben-Or's technique~\cite{benor1983lower} works as follows: we formulate a problem as a question \enquote{\(x \in W\)?} for some set \(W \subseteq \R^d\) and then bound the height of any algebraic computation tree deciding this membership question by bounding the number of connected components of \(W\). Ben-Or~\cite{benor1983lower} improved known lower bounds, e.g., for the knapsack problem~\cite{dobkin1978lower}, and established new ones for a variety of problems including element distinctness and geometric constructions with ruler and compass. Sacristán~\cite{sacristan1999lower} summarizes the results related to Ben-Or's technique. 

\section{An Upper Bound} \label{sec::upperbound}

We divide the sorted arrays \(A_1\), \(A_2\), \dots, \(A_k\) into groups of size~\(s\). Then, we merge the \(s\) arrays in the \(j\)-th group into one sorted array \(B_j\), e.g., by maintaining a min-heap of size \(s\) storing for each array the smallest entry that has yet to be inserted into \(B_j\). Finally, we apply fractional cascading~\cite{chazelle1986fractional} on \(B_1\),~\dots, \(B_{\ceil*{k/s}}\). This construction takes \(O(n \log s)\) time and occupies \(O(n)\) space. We answer a predecessor query for \(q\) in \(O(k/s + \log n)\) time by determining the predecessors \(p_1,\dots, p_{\ceil*{k/s}}\) of \(q\) in each \(B_1\), \(B_2\), \dots, \(B_{\ceil*{k/s}}\), respectively; the largest of these values is the predecessor of \(q\) in \(A\). When \(k = \omega(\log n)\), we obtain a query time of \(O(\log n)\) and a  construction time of \(O(n \log(\frac{k}{\log n}))\) by choosing \(s = \Theta\left({k/\log n}\right)\).  

\section{A Lower Bound} \label{sec::lowerbound}

Our general approach is as follows. Let \(T(n,m)\) be the total time required for answering a sequence of \(m\) queries. Assume we have a data structure with construction time \(C(n)\) supporting queries in \(Q(n)\) time. Answering \(m = \floor*{n / Q(n)}\) queries takes \(T(n,m) \le m Q(n) + C(n) \le n + C(n) \) time. Therefore, any lower bound of \(T(n,\floor*{n / Q(n)}) = \Omega(X)\), with \(X = \omega(n)\), implies a lower bound of \(C(n) = \Omega(X)\). We shall use Ben-Or's technique to find a suitable \(X\). 

Consider the following \emph{batch verification} variant of strict predecessor search: We are given \(k\) sorted arrays \(A_1\), \(A_2\), \dots, \(A_k\) of lengths \(n_1,n_2,\dots,n_k\), respectively, and we are given \(m\) query points \(q_1,q_2,\dots, q_m \in \R \) alongside with \(m\) supposed answers \(p_1,p_2,\dots,p_m\in \R\). We would like to check whether \(p_i\) is indeed the strict predecessor of \(q_i\) among all values in \(A_1,A_2,\dots, A_k\) for all \(i=1,2,\dots,m\). This batch verification problem corresponds to the membership problem for 
\begin{align*}
	W_m &\coloneqq \Set*{ \begin{array}{c} A_1 \in \R^{n_1}  \\ \vdots \\ A_k \in \R^{n_k} \\ q \in \R^m \\  p \in \R^m \end{array} }{
	\begin{array}{l} \text{The entries of } A_1, \dots, A_k \text{ are sorted and} \\
	 p_i \text{ is the strict predecessor of query } q_i  \\
	\text{among } A_1, \dots, A_k  \text{ for all } i = 1, \dots, m .\end{array}} \subset \R^{n + 2m} \enspace . 
\end{align*}

According to Ben-Or's theorem, deciding the membership problem \(W_m\) for the batch verification problem takes \(\Omega(\log\#W_m - d)\) time, where \(\#W_m\) is the number of components of \(W_m\) and \(d = n + 2m\) is the dimension of \(W_m\). As the next step, we establish a lower bound on \(\#W_m\) by identifying a certain number of points that belong to pairwise distinct connected components of \(W_m\).

To study the structure of \(W_m\), we start with some instance \(x \in W_m\) of batch verification, i.e., a point \(x \in \R^{n+2m}\) encoding \(k\) sorted arrays \(A_1,A_2,\dots, A_k\), queries \(q_1,q_2,\dots,q_m\), and answers \(p_1,p_2,\dots,p_m\) such that \(p_i\) is the strict predecessor of \(q_i\) among all array entries for \(i=1,2,\dots,m\). We can continuously move some of the entries of \(x\) without leaving \(W_m\). For instance, we remain in \(W_m\) when moving a query without changing its strict predecessor. Other changes, like moving the supposed answer \(p_i\) for query \(q_i\) without moving the corresponding array entry, cause us to leave \(W_m\). The components of \(W_m\) consist of instances that can reach one another via a continuous deformation without leaving \(W_m\). 
\vspace{-\baselineskip}

\begin{figure}[htb]
	\centering 
	\includegraphics[scale=1,page=6]{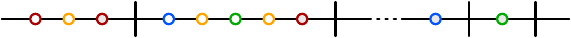}
	\caption{A distribution of array entries (empty circles) and queries (vertical bars). The colors indicate the array containing an entry, i.e., all entries of color \(i\) belong to \(A_i\).
	\vspace{-3\baselineskip}
	\label{fig::order}}
\end{figure}
\begin{figure}[htb]
	\centering
	
	\subfloat[We can swap array entries that are no strict predecessors.]{\includegraphics[page=2]{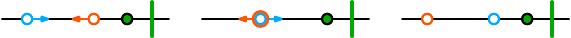}}
	
	\subfloat[We can swap predecessors entries with non-predecessor entries.]{\includegraphics[page=3]{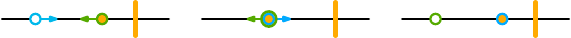}}
	
	\subfloat[We \emph{cannot} move an array entry through a query.]{\includegraphics[page=4]{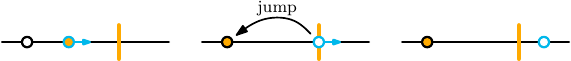}}
	
	\caption{Legal and illegal changes to the order of queries (vertical bars), array entries (empty circles), and strict predecessors (color of circle centers). We can swap entries and queries as shown in (a) and (b) without leaving \(W\). As depicted in (c), moving an array entry through a query point \(q\) (or vice versa) froces us out of \(W\), as the strict predecessor of \(q\) to would have to discontinuously jump to a new position. \label{fig::swaps}}
	\vspace{-\baselineskip}
\end{figure} 

Consider the order of the array entries, query points, and answers of an instance \(x \in W_m\), as illustrated in Figure\nobreakspace \ref {fig::order}. We estimate \(\#W_m\) by counting orders in separate components of \(W_m\).  Figure\nobreakspace \ref {fig::swaps} summarizes which changes lead to the same component and which changes leave the component. Most importantly, we cannot move a query value through an array entry or vice versa without causing the corresponding answer to discontinuously \emph{jump} to a new position. 

To count the components of \(W_m\), we consider the different ways to distribute \(n\) distinct values \(x_1, x_2, \dots, x_n\) into sorted arrays \(A_1,A_2,\dots, A_k\) and then we trap these distributions in as many separate components of \(W_m\) as possible by placing query values. There are \(\frac{n!}{n_1! n_2! \cdots n_k!}\) ways to distribute \(n\) distinct values into \(k\) sorted arrays of sizes \(n_1,n_2,\dots, n_k\).
\begin{figure}[ht]
	\centering \vspace{-\baselineskip}
	\includegraphics[scale=1,page=8]{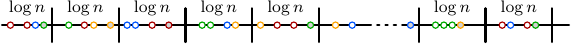}
	\caption{Trapping distributions of the values (circles) to the arrays (colors) using only \(\frac{n}{\log n}\) queries (vertical bars). We place one query point every \(\log n\) entry values. Using the legal swaps from Figure\nobreakspace \ref {fig::swaps}, we can reorder the entries between two queries. \label{fig::trapmany}} \vspace{-\baselineskip}
\end{figure}

As illustrated in Figure\nobreakspace \ref {fig::trapmany}, we place one query point every \(\log n\) array entries. Since we can swap any two of the \(\log n\) entries between two queries, the number of components shrinks by a factor of at most \((\log n)!\) per query compared to placing one query between every two array entries, i.e.,
\[
	\#W_{\frac{n}{\log n}} \ge \dfrac{W_{n-1}}{(\log n)!^{\frac{n}{\log n}}} \ge \dfrac{n!}{n_1! n_2! \cdots n_k! \cdot (\log n)!^{\frac{n}{\log n}}}  = \Omega\left(2^{n \log( \frac{k}{\log n})}\right) \enspace,
\] 
where the asymptotic bound follows from Stirling's formula and from the fact that the expression is maximized when the lengths of the arrays are balanced.
\begin{theorem} Consider \(k\) sorted arrays \(A_1,A_2,\dots, A_k\) containing \(n\) entries in total, and let \(A\) be the sorted array that results from merging \(A_1,A_2,\dots, A_k\). 
When \(k = \omega(\log n)\), every data structure that supports strict predecessor queries in \(A\) with a query time of \(O(\log n)\) requires \(  \Omega (n \log( \frac{k}{\log n}) ) \) construction time. 
\end{theorem}

\section{Future Work} \label{sec::conclusion}

In future research, we shall attempt to reestablish our lower bound for non-strict predecessor queries, e.g., by augmenting the algebraic computation tree model with support for symbolic perturbation~\cite{emiris1995general}. Moreover, we shall apply our approach to derive new lower bounds for the construction of other data structures.

\bibliography{pred_lower_bound}{}

\begin{thebibliography}{1}
\providecommand{\url}[1]{\texttt{#1}}
\providecommand{\urlprefix}{URL }

\bibitem{benor1983lower}
Ben-Or, M.: Lower bounds for algebraic computation trees. In: Proceedings of
  the Fifteenth Annual ACM Symposium on Theory of Computing. pp. 80--86 (1983)

\bibitem{chazelle1986fractional}
Chazelle, B., Guibas, L.J.: Fractional cascading: I. {A} data structuring
  technique. Algorithmica  1(2),  133--162 (1986)

\bibitem{dobkin1978lower}
Dobkin, D.P., Lipton, R.J.: A lower bound of {\({\frac 1 2 n^2}\)} on linear
  search programs for the knapsack problem. J. Comput. Syst. Sci.  16(3),
  413--417 (1978)

\bibitem{emiris1995general}
Emiris, I.Z., Canny, J.F.: A general approach to removing degeneracies. {SIAM}
  Journal on Computing  24(3),  650--664 (1995)

\bibitem{sacristan1999lower}
Sacristán, V.: Lower bounds for some geometric problems. Tech. Rep.
  MA2-IR-98-0034, Universitat Politècnica de Catalunya (1999)

\bibitem{sen2008lower}
Sen, P., Venkatesh, S.: Lower bounds for predecessor searching in the cell
  probe model. J. Comput. Syst. Sci.  74(3),  364--385 (2008)

\end{thebibliography}
\bibliographystyle{splncs03}

\end{document}